\documentclass[a4paper,11pt]{article}
\usepackage{pos}

\title{AI-assisted design of experiments at the frontiers of computation: methods and new perspectives}

\author*[a]{Pietro Vischia}

\affiliation[a]{Universidad de Oviedo and ICTEA,\\
  Calle San Francisco 3, 33007 Oviedo, Principado de Asturias, España}

\emailAdd{pietro.vischia@cern.ch}

\abstract{Designing the next generation colliders and detectors involves solving optimization problems in high-dimensional spaces where the optimal solutions may nest in regions that even a team of expert humans would not explore.

Resorting to Artificial Intelligence to assist the experimental design introduces however significant computational challenges in terms of generation and processing of the data required to perform such optimizations: from the software point of view, differentiable programming makes the exploration of such spaces with gradient descent feasible; from the hardware point of view, the complexity of the resulting models and their optimization is prohibitive. To scale up to the complexity of the typical HEP collider experiment, a change in paradigma is required.

In this talk I will describe the first proofs-of-concept of gradient-based optimization of experimental design and implementations in neuromorphic hardware architectures, paving the way to more complex challenges.
}

\FullConference{42nd International Conference on High Energy Physics (ICHEP2024)\\
18-24 July 2024\\
Prague, Czech Republic\\}

\begin{document}
\maketitle

\section{Introduction}
\label{sec:intro}
Detectors and accelerators that characterize modern particle physics experiments are complex apparata, whose construction can span a decade and whose cost requires international cooperation to make the construction and operation of the experiment feasible. The 2020 European Strategy (EUSUPP) strategy recommends the implementation of "New large, long-term projects, pushing technological skills to the limit"~\cite{CERN-ESU-015}: in the current worldwide landscape of scarse resources, political instability, and environmental concerns, it is imperative for us scientists to optimize our experimental designs to extract maximal scientific value within constraints. Furthermore, performing a constrained optimization of an experiment design can include not only constraints coming from the available resources, but also from any sort of limitation (e.g. geometrical positioning of detection elements, cables, etc.).

Constrained design optimization is not a novel concept (it is widely used in industry), but its application to particle physics is novel: the data processed by instruments based on the interaction between light and matter are instrinsically stochastic: the usual optimization paradigm based on likelihood maximization therefore fails, because the likelihood is intractable. Costly Monte Carlo simulators are consequently needed to generate these data, and a generic pipeline can be described as follows:
\begin{enumerate}
    \item A multidimensional stochastic variable $x\sim f(x, \phi)$ is sampled depending on parameters $\phi$;
    \item Data interact with the detector, creating readouts $z\sim p(z | x(\phi), \theta)$, where $\theta$ are construction parameters of the detector. The readouts $z$ depend implicitly on the parameters $\phi$ through $x$;
    \item (optional) High-level observables are build based on the readouts: $\zeta(\phi, \theta) = R[z(\phi, \theta), \nu(\phi)]$, where $\nu$ are nuisance parameters representing systematic uncertainties in the measurement;
    \item A low-dimensional summary statistic is built, $s = A[\zeta(\phi, \theta)]$, and used for inference.
\end{enumerate}
Intractable likelihood problems must be solved via likelihood-free methods: histograms are the basic likelihood-free method that has been in the standard practice in particle physics for decades, binning the data and assuming the full likelihood is a product of Poisson distributions~\cite{VISCHIA2020100046}. 
Inference can be performed but not optimised through histograms. Because of these limitations, experiment design optimization in particle physics used to be performed by generating a limited amount of data and a very limited amount of simulated experiment settings for a small number of design parameters and a few values for each of these parameters. The interpolation capabilities of modern Artificial Intelligence (AI) algorithms can be exploited to achieve the optimization of vast parameter spaces without having to generate a prohibitive amount of simulated data and configurations.

\section{Differentiable software}

Simulated particle physics data contain the full information of the true value of the parameters used to generate them. Using a supervised learning setting is therefore natural. Supervised learning is therefore the preferred setting to make design optimization studies based on simulated data: the problem of learning formally becomes a problem of interpolation with the additional requirement that the solutions should generalize to unseen data or simulated data, and the solution is formalised as empirical risk minimization (ERM)~\cite{vapnik} solved by gradient descent~\cite{vidal2017mathematicsdeeplearning}. The empirical loss, $\mathcal{L}(\theta)$, is computed comparing the estimated and true output of the model: the gradient of the empirical loss is then computed, and propagated through the model in cascade, using the chain rule, to update the model parameters until the loss is minimised. Gradients are computed with automatic differentiation (AD)~\cite{walther}, a technique, based on the decomposition of computer code into direct acyclic graphs, to evaluate exact derivatives with limited computational overhead.

Classical artificial neural networks (ANNs) are parametric models based on the connectivity of simplified mathematical representations of neurons, the perceptrons~\cite{perceptron}, resulting in models whose computational efficiency depend on recasting the problem as matrix multiplication: in hardware, this involves the use of chips such as CPUs, GPUs, and FPGAs. In Section~\ref{sec:paradigms} I will discuss two changes of paradigm that will vastly improve the computational characteristics of neural networks.

The optimization of parametric models via AD can be generalised from the weights of a ANN to any kind of parameter of a given model, enabling AI-assisted optimization of anything parameterizable in computer code: this paradigm, popularized by LeCun in 2018~\cite{yannlecun}, but originally proposed by Schmidhuber in 1990~\cite{schmidhuber1990making}, gave rise to a growing literature. Because of space constraints, in the following I will focus on a few examples from my work and that of my collaborators.

\section{Inference}
Inference in particle physics proceeds mostly through binned or unbinned profile likelihood ratio estimation~\cite{VISCHIA2020100046}: AI has been used since decades to improve the quality of classifiers or regressors, to increase the separation of the probability distribution function of the physics hypotheses that are being tested one against the other, mostly resulting in learning differentiable surrogates of likelihood ratios. Recently, some authors claimed~\cite{theoreticiansthinkingtheyknowthings} to have invented this method, but a basic literature search disproves their claim~\cite{PhysRevD.87.052009,CMS-PAS-HIG-16-004,vischialol}, showing that ML discriminants (binned or unbinned) for inference are routinely used since decades and that, in particular, learning likelihood ratios is explicitly present in literature (in both binned and unbinned form) since more than five years.

\section{Experiment Design}
Differentiable programming for experiment design consists in optimising the pipeline described in Section~\ref{sec:intro} with respect to the design parameters of the experiment, using gradient descent. The goal is to optimize solutions in a high-dimensional, nontrivial parameter space where the sheer complexity and asymmetries of the problem would defy human attempts. The loss function can be decomposed into a physics output component and a cost component that can be arbitrarily penalized~\cite{whitepaper}, and the minimization problem is:
\begin{equation}
    \hat{\theta} = arg~min_{\theta} \int L[A(\zeta(\phi,\theta),c(\theta)] p(z | x(\phi),\theta) f(x,\phi) dx dz\;,
\end{equation}
where $p(z | x(\phi),\theta)f(x,\phi)$ must be writeable in a closed form, or substituted with a differentiable surrogate model (e.g. a neural network).
The aim of the optimization process is, however, not to give \textit{the} optimal solution and mandate its use. The aim is rather to assist the physicist with a landscape of solutions around the minimum of the loss function: the physicist then will use her judgment and domain knowledge to choose the most feasible solution, integrating in the decision any consideration that was not parameterized in the model, either because of discreteness (discrete problems are in general nondifferentiable) or because of simplifications or other needs. The proposed range of optimal parameters will depend on the steepness and convexity characteristics of the loss function landscape around the minimum: smoothing techniques may be needed, as well as a thorough investigation of the structure of minima (see e.g. Ref.~\cite{PhysRevLett.127.278301} or the works by Belkin, for an overview).

\section{Proofs of Concept}
\label{sec:proofsofconcept}
The holy grail is the optimization of experiments of the scale of the Large Hadron Collider (LHC). To achieve this ambitious goal, several efforts focussed on smaller-scale experiments, studying diverse settings and learning important lessons transferrable to the large scale.

\begin{figure}
    \centering
    \includegraphics[width=0.5\linewidth]{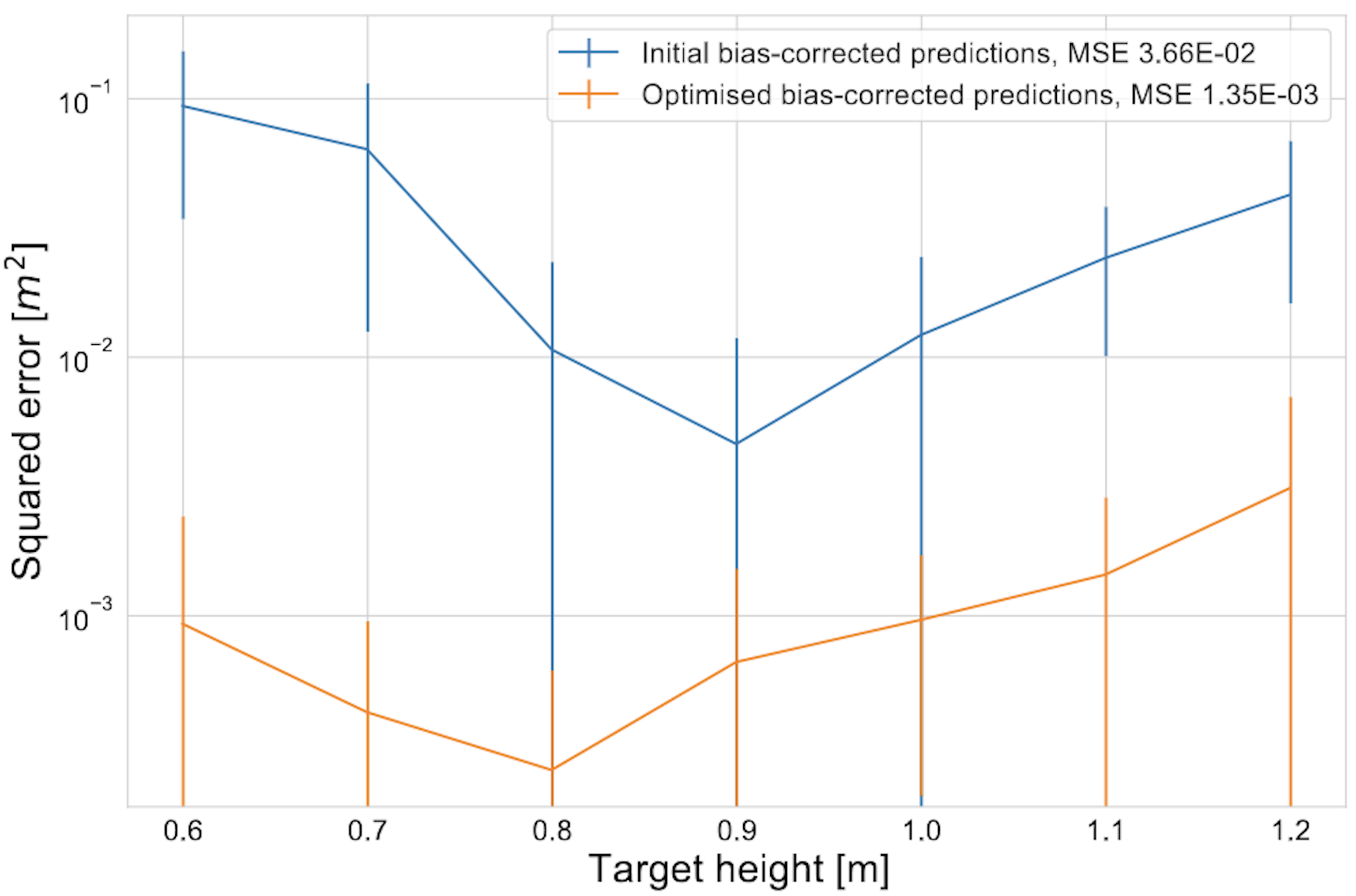}
    \caption{Mean square error of the bias-corrected predictions before and after the optimization loop. Reproduced with authorization from Ref.~\cite{Strong_2024}.}
    \label{fig:tomopt}
\end{figure}

The first proof of concept of AD-assisted experiment design therefore focussed on the optimization of a simple muon tomography setup. Published with the pen name \textsc{TomOpt}~\cite{Strong_2024}, the work demonstrates (and provides open source code for) the optimization of the placement and size of a number of muon detector panels around a ladle furnace to infer the fill level of molten steel inside. Visual inspection cannot help, because of floating impurities. The steel level is inferred from a three-dimensional voxelized radiation length map of the material in the ladle: prediction biases coming from ignoring multiple scatterings are corrected for. The promising results are illustrated in Fig.~\ref{fig:tomopt}, where the mean square error significantly improves after the optimization of the experiment layout. Further results have been obtained with different constraints on the detector panels (accepted by the MARESEC conference, to be published).

\begin{figure}
    \centering
    \includegraphics[width=0.7\linewidth]{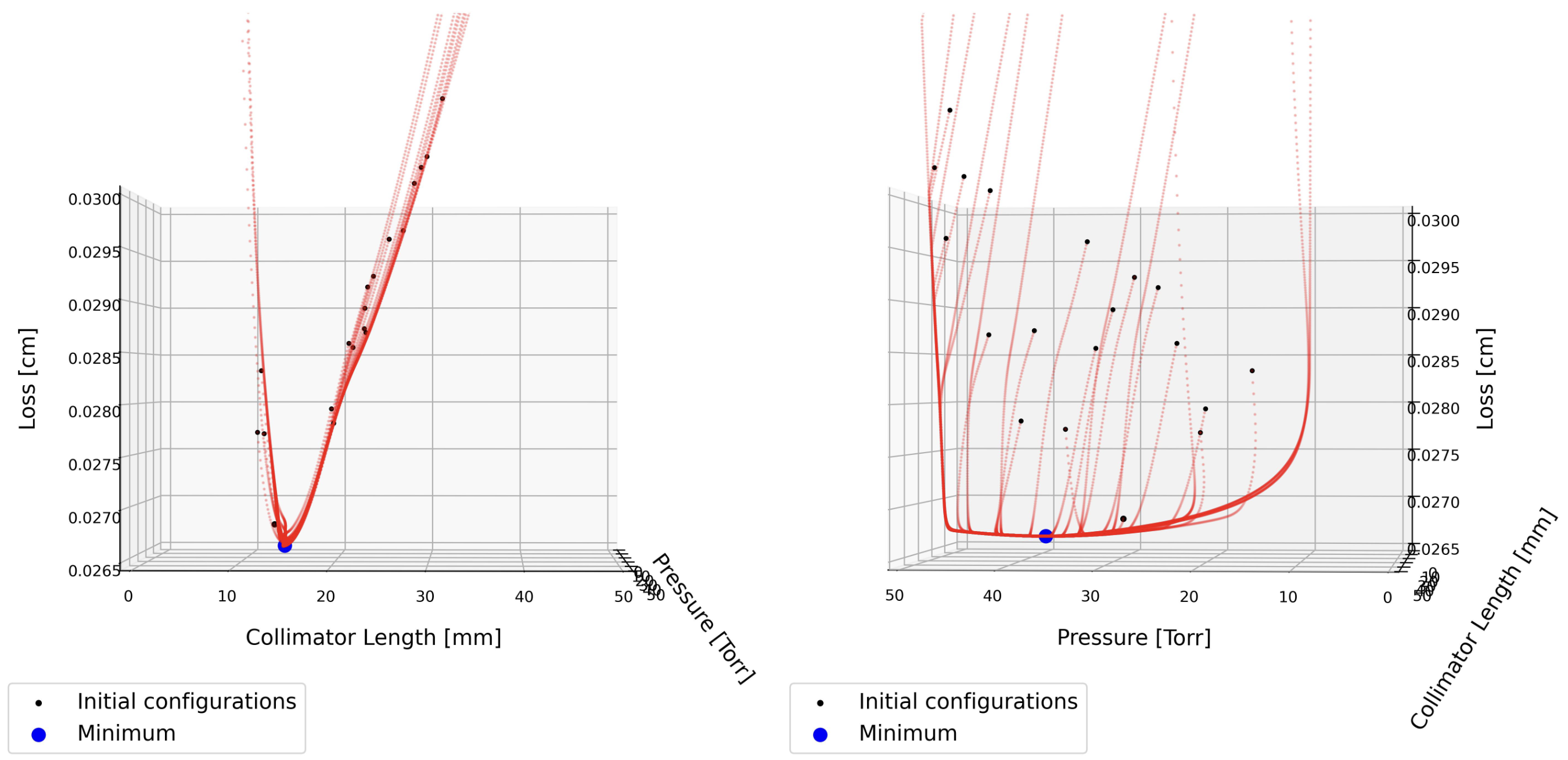}
    \caption{Mean square error of the bias-corrected predictions before and after the optimization loop. Reproduced with authorization from Ref.~\cite{mariapereira}.}
    \label{fig:oppac}
\end{figure}

An impressive set of results has been obtained in the optimization of a parallel-plate avalanche counter with optical readout in the context of a neutron tomography experiment~\cite{mariapereira}. The optimised solutions for the two design parameters of interest remarkably converges to the same solution regardless of the starting point of the optimization, as illustrated in Fig.~\ref{fig:oppac}, and the optimization for one of the parameters coincides exactly with the results from independent studies via brute force scan of configurations. Similar behaviours have been observed in optimization studies for the layout of a gamma ray observatory~\cite{dorigo2023endtoendoptimizationlayoutgamma}.

\section{New computational paradigms}
\label{sec:paradigms}
The proofs of concept described in Section~\ref{sec:proofsofconcept} are computational challenging even in these low-dimensional cases. For instance, computational tricks (such as dedicated CUDA kernels) had to be devised to make the optimization in \textsc{TomOpt} feasible for suitable desired resolutions. Scaling up to experiment of the complexity of the LHC (and of its successors) is unfeasible with the current technology based on CPUs, GPUs, FPGAs. A change in paradigm is therefore needed.

\subsection{Neuromorphic Computing}
The first change of paradigm, that will be implementable in the medium term, is to replace the matrix-multiplication-based ANNs made of perceptrons with event-based computation in spiking neural networks (SNNs). These models are inspired by the workings of biological neurons and are based on the generation of spikes in the membrane potential: the dynamics of these spikes can be approximated by electrical circuits solvable with partial differential equations. Ref.~\cite{bolita} demonstrates that these neurons can mimic biological neuronal dynamics to a good extent: however, solving PDEs in a large scale is still computationally challenging.  The leaky integrate-and-fire (LIF) models vastly simplify the equivalent circuit, making artificial SNNs feasible and constituting a paradigm shift in information encoding: hardware implementations of SNNs exploit time encoding to vastly reduce the need of memory access in sparse problems, resulting in chips with extremely low power consumption~\cite{neurom}.
Neuromorphic computing is particularly attractive when the data have an intrinsic time-dependent encoding. Recent efforts are e.g. proposing neuromorphic readout for granular calorimeters~\cite{enricolupi}. I hereby propose a neuromorphic readout for the Q-Pix detector~\cite{PhysRevD.106.032011}: Q-Pix output is based on time-dependent pulses: a neuromorphic readout would make the data immediately processable without the need of any intermediate encoding or reconstruction.

\subsection{Quantum Computing}
Quantum Computing is based on the representation of data as qubits (\textit{quantum bits}). Instead of the binary representation characteristic of classical bits, qubits represent of data as quantum states on a sphere (Bloch sphere), therefore assuming infinite values. Representing data operations as quantum operators (gates) acting on qubits makes it possible to perform computations "on the whole distributions", obtaining specific output values only when the wave functions collapse as a result of \textit{measurement} operations~\cite{schuld}. Representing data and machine learning algorithms via qubits and quantum circuits makes it therefore possible to obtain the same accuracy as classical algorithms but using orders of magnitude less data~\cite{terashi}. A recent demonstration has been performed in the context of Quantum Natural Language Processing~\cite{manueluria}
Quantum machine learning (QML) has another very desirable characteristic that makes it perfect for experiment design: neural networks can be naturally represented by unitary and trace operators, and the resulting quantum circuits are analytically differentiable. This, coupled with the speedup given by the \textit{measure-at-the-end} paradigm, makes QML a natural candidate for optimizing complex experiments, in the long term.

\section{Conclusions}
Designing the next generation colliders and detectors involves solving optimization problems in high-dimensional spaces where the optimal solutions may nest in regions that even a team of expert humans would not explore.
In this manuscript, I have described the first proofs of concept of gradient-based optimization of experimental design and paved the way towards more complex challenges. I have proposed the implementation of spiking neural networks for Q-Pix apparata, and foreseen the implementation of AI-assisted experiment design with QML techniques.

\end{document}